\documentclass[12pt,preprint]{aastex63}
\usepackage{natbib}

\shorttitle{H$\alpha$ Emission in HR~6819} 
\shortauthors{Gies et al.} 
\input{epsf} 

\begin{document} 
\accepted{2020 July 11, ApJL, in press} 
 
\title{The H$\alpha$ Emission Line Variations of HR~6819} 

\correspondingauthor{Douglas Gies}
\email{gies@chara.gsu.edu, lwang@chara.gsu.edu}

\author[0000-0001-8537-3583]{Douglas R. Gies}
\affiliation{Center for High Angular Resolution Astronomy and
 Department of Physics and Astronomy,\\
 Georgia State University, P. O. Box 5060, Atlanta, GA 30302-5060, USA}

\author[0000-0003-4511-6800]{Luqian Wang}
\affil{Center for High Angular Resolution Astronomy and
 Department of Physics and Astronomy,\\
 Georgia State University, P. O. Box 5060, Atlanta, GA 30302-5060, USA}

 
 
\begin{abstract} 
The star system HR~6819 was recently proposed by Rivinius et al.\ 
as the site of the nearest example of a stellar mass black hole. 
Their spectra show evidence of two components: a B3~III star
in a 40~d orbit and a stationary B-emission line star.  Based 
upon the orbital mass function and the lack of evidence of a
spectral component with reflex orbital motion, Rivinius et al.\ 
suggested that HR~6819 is a triple system consisting of an 
inner B3~III star plus black hole binary orbited by a distant
Be star. Here we present an alternative model based upon an 
examination of the H$\alpha$ emission line in their spectra. 
We show that the emission line displays the small reflex motion 
expected for the companion in the inner orbit and shows variations 
in profile shape related to orbital phase.  This indicates 
that HR~6819 is a binary system consisting of a massive Be
star and a low mass companion that is the stripped down 
remnant of a former mass donor star in a mass transfer binary.   
\end{abstract} 
 
 
 
\section{Introduction}                              

\citet{Rivinius2020} recently presented a spectroscopic investgation
of the bright, early-type multiple system HR~6819 (HD~167128, QV~Tel).
Their analysis shows that the spectrum is dominated by three 
spectral components: a narrow-lined B3~III star in a 40~d orbit, 
a stationary Be star, and continuum and emission line flux 
from the disk surrounding the Be star.  The orbital mass function 
and probable mass of the B3~III star indicate that the companion
is massive ($> 4.2 M_\odot$), yet its expected spectral features 
are absent from the spectrum.  This led \citet{Rivinius2020} to
suggest that the companion is a non-accreting and X-ray quiet 
black hole and that the Be star is a distant tertiary component. 

There is growing evidence that many Be stars are the product 
of mass transfer in an interacting binary \citep{Pols1991}. 
The initially more massive component in a close binary is the 
first to expand after core hydrogen exhaustion, and as it 
reaches the dimensions of the critical Roche surface, it 
will begin to transfer mass and angular momentum to the 
companion star.  If the system avoids a merger, the donor 
star will end up as a low mass core stripped of its envelope, 
and the mass gainer will appear as a rejuvenated, fast rotating, 
main sequence star.  Be stars are fast rotating objects that 
are shedding angular momentum into an outflowing decretion 
disk \citep{Rivinius2013}, so they are potential candidates of
systems spun up by binary mass transfer.  The stripped companions 
are generally faint and difficult to detect, but 
spectroscopic searches in the ultraviolet part of the spectrum 
have begun to reveal the spectral lines of the hot, low mass companions 
\citep{Wang2018}.

The presence of a Be star component in the spectrum of HR~6819 
suggests another interpretation of the system. 
It is possible that the B3~III stellar component is actually 
a low-mass, stripped down star that is still relatively 
young and luminous.  In this case, the Be star would be the 
companion in the 40~d binary instead of a black hole, and it 
should exhibit the orbital reflex motion around the B3~III star.
Here we present an analysis of the H$\alpha$ emission line that
is formed in the disk surrounding the Be star.  We show that 
the radial velocities associated with the disk do indeed 
display a small reflex orbital motion (Section 2) as well 
as profile shape variations that are related to the orbital 
phase of the inner binary (Section 3).  These results imply 
that the system is a binary that consists of a rapidly 
rotating Be star and a low-mass, bright companion that is 
the core of the former mass donor (Section 4).   
 
 
\section{H$\alpha$ Wing Bisector Velocities}        

The hydrogen Balmer emission lines form in a Keplerian rotating 
disk surrounding the Be star, and the profile generally appears 
as a double-peaked profile \citep{Sigut2020}.  The fastest moving 
gas is found closest to the star, so the extreme wings of the 
emission line provide the best proxy of the motion of the 
underlying star.  The H$\alpha$ emission line
is the strongest emission line in the spectrum of HR~6819, 
so it is the ideal feature to search for evidence of disk 
velocity variations associated with the orbital motion if
the Be star is part of the 40~d binary.

We downloaded the full set of FEROS spectra from \citet{Rivinius2020}
for their Set~A group from the ESO 1.5~m 
telescope\footnote{https://www.lsw.uni-heidelberg.de/projects/instrumentation/Feros/ferosDB/search.html} 
(12 spectra) and their Set~B group from the ESO/MPG 2.2~m 
telescope\footnote{http://archive.eso.org/wdb/wdb/adp/phase3\_main/form?collection\_name=FEROS} (51 spectra).  
The spectra were rebinned onto a standard wavelength grid 
in increments of $\log \lambda$ for a reduced spectral 
resolving power of $R = 25000$ (to increase S/N), and they 
were rectified to a unit continuum by a linear fit to line-free 
regions over the wavelength range from 6500 to 6700 \AA . 
This spectral window records the H$\alpha$ $\lambda 6563$ 
emission line plus the photospheric absorption features of 
\ion{C}{2} $\lambda\lambda 6578, 6583$ and \ion{He}{1} $\lambda 6678$.

The first step was to confirm the orbital elements of \citet{Rivinius2020}
by measuring radial velocities for the narrow-lined B3~III star 
(denoted by star 1).  We formed cross-correlation functions (CCFs) of 
the observed spectra and a model spectrum over the range covering 
the \ion{C}{2} and \ion{He}{1} features.  The model spectrum was 
derived from the BLUERED spectral grid \citep{Bertone2008} for 
solar abundances, $T_{\rm eff} =17100$~K, $\log g = 3.5$, and no
rotational broadening \citep{Rivinius2020}.  The radial velocities 
measured from the CCF peaks are presented in Table~1 that lists the 
heliocentric Julian date of mid-exposure, the orbital phase, 
and the radial velocities for star 1 (plus wing bisector velocities
for the Be star; see below).  We also include a single 
measurement from the far ultraviolet spectrum obtained with 
the {\it International Ultraviolet Explorer (IUE)} (SWP31220) that we measured 
using the method described by \citet{Wang2018}.  The orbital elements 
were then fit using the non-linear, least squares program of 
\citet{Morbey1974}, and the resulting orbital elements are listed
in Table~2.  These elements are in good agreement with those presented
by \citet{Rivinius2020}, and in particular, the {\it IUE} measurement 
from 1987 confirms the orbital period they derived.  The orbital 
radial velocity curve is shown in Figure~1 with phase zero defined 
by the epoch of periastron given in Table~2. 
 
\placetable{tab1}      
\begin{deluxetable}{lcrrr} 
\tabletypesize{\scriptsize} 
\tablewidth{0pt} 
\tablenum{1} 
\tablecaption{Radial Velocity Measurements \label{tab1}} 
\tablehead{ 
\colhead{Date}              & 
\colhead{Orbital}           & 
\colhead{$V_r(1)$}          & 
\colhead{$V_r(2)$}          & 
\colhead{$V_r^{sub}(2)$}    \\ 
\colhead{(HJD$-$2,400,000)} & 
\colhead{Phase}             & 
\colhead{(km s$^{-1}$)}     & 
\colhead{(km s$^{-1}$)}     & 
\colhead{(km s$^{-1}$)}     } 
\startdata 
 46967.186 &  0.029 &     2.1 & \nodata & \nodata \\
 51373.677 &  0.280 &   -41.8 &    17.1 &    13.8 \\
 51374.675 &  0.305 &   -48.9 &    15.4 &    12.3 \\
 51375.653 &  0.329 &   -44.2 &    15.3 &    12.4 \\
 51376.724 &  0.355 &   -28.4 &    13.3 &    10.8 \\
 51378.682 &  0.404 &   -30.6 &    13.4 &    11.5 \\
 51380.824 &  0.457 &   -16.0 &    11.8 &    10.8 \\
 51383.673 &  0.528 &    11.3 &     9.0 &     9.4 \\
 51384.743 &  0.554 &    22.2 &     7.3 &     8.3 \\
 51385.760 &  0.579 &    27.1 &     6.9 &     8.2 \\
 51390.745 &  0.703 &    61.8 &     3.6 &     6.7 \\
 51393.627 &  0.774 &    70.7 &     2.4 &     5.8 \\
 51394.758 &  0.802 &    66.4 &     1.6 &     5.1 \\
 53138.848 &  0.044 &   -10.4 &     6.1 &     5.7 \\
 53139.683 &  0.065 &   -16.2 &     6.7 &     5.8 \\
 53143.808 &  0.167 &   -52.7 &    11.2 &     8.3 \\
 53144.754 &  0.190 &   -56.3 &    13.6 &    10.3 \\
 53149.763 &  0.314 &   -53.4 &    13.7 &    10.5 \\
 53149.766 &  0.315 &   -53.4 &    13.5 &    10.4 \\
 53154.669 &  0.436 &   -21.9 &     8.9 &     7.7 \\
 53159.805 &  0.563 &    18.0 &     2.8 &     4.2 \\
 53160.687 &  0.585 &    27.9 &     4.3 &     6.1 \\
 53162.863 &  0.639 &    40.2 &    -1.9 &     1.2 \\
 53183.549 &  0.152 &   -47.6 &    12.3 &     9.5 \\
 53185.551 &  0.202 &   -53.7 &    13.5 &    10.2 \\
 53187.667 &  0.254 &   -55.0 &    14.3 &    10.8 \\
 53188.688 &  0.280 &   -53.0 &    11.8 &     8.6 \\
 53190.705 &  0.330 &   -51.8 &    12.8 &     9.9 \\
 53194.692 &  0.428 &   -29.4 &     9.3 &     7.9 \\
 53195.564 &  0.450 &   -17.3 &     7.1 &     6.2 \\
 53196.551 &  0.474 &    -5.0 &     7.7 &     7.2 \\
 53197.670 &  0.502 &     7.0 &     6.9 &     7.0 \\
\enddata 
\end{deluxetable} 
 
\pagebreak
 
\placetable{tab1}      
\begin{deluxetable}{lcrrr} 
\tabletypesize{\scriptsize} 
\tablewidth{0pt} 
\tablenum{1} 
\tablecaption{Radial Velocity Measurements ({\it continued}) \label{tab1}} 
\tablehead{ 
\colhead{Date}              & 
\colhead{Orbital}           & 
\colhead{$V_r(1)$}          & 
\colhead{$V_r(2)$}          & 
\colhead{$V_r^{sub}(2)$}    \\ 
\colhead{(HJD$-$2,400,000)} & 
\colhead{Phase}             & 
\colhead{(km s$^{-1}$)}     & 
\colhead{(km s$^{-1}$)}     & 
\colhead{(km s$^{-1}$)}     } 
\startdata 
 53199.571 &  0.549 &    26.0 &     4.7 &     5.8 \\
 53202.646 &  0.626 &    48.7 &     1.7 &     4.2 \\
 53204.513 &  0.672 &    64.2 &     0.7 &     4.0 \\
 53205.697 &  0.701 &    65.3 &    -1.0 &     2.8 \\
 53207.502 &  0.746 &    71.5 &    -2.6 &     1.4 \\
 53226.597 &  0.219 &   -55.1 &    15.8 &    12.2 \\
 53230.491 &  0.316 &   -45.5 &    13.4 &    10.4 \\
 53231.570 &  0.343 &   -42.0 &    13.1 &    10.2 \\
 53232.503 &  0.366 &   -42.5 &    13.0 &    10.3 \\
 53239.495 &  0.539 &    17.2 &     6.2 &     7.0 \\
 53240.603 &  0.567 &    27.4 &     4.8 &     6.2 \\
 53243.516 &  0.639 &    53.9 &     2.4 &     5.2 \\
 53244.552 &  0.665 &    63.1 &     1.0 &     4.1 \\
 53245.511 &  0.688 &    63.8 &     1.2 &     4.4 \\
 53246.620 &  0.716 &    61.7 &     0.8 &     4.6 \\
 53247.553 &  0.739 &    64.5 &     0.3 &     4.2 \\
 53248.523 &  0.763 &    66.0 &     0.9 &     4.9 \\
 53254.540 &  0.912 &    47.2 &     2.0 &     4.8 \\
 53255.516 &  0.936 &    32.8 &     3.5 &     5.6 \\
 53256.556 &  0.962 &    26.2 &     5.7 &     7.2 \\
 53258.493 &  0.010 &    13.7 &     8.0 &     8.2 \\
 53259.532 &  0.036 &     0.8 &     9.6 &     9.2 \\
 53260.537 &  0.061 &    -8.3 &     9.8 &     8.9 \\
 53261.542 &  0.086 &   -16.5 &    12.1 &    10.5 \\
 53262.530 &  0.110 &   -22.1 &    13.7 &    11.6 \\
 53263.537 &  0.135 &   -39.8 &    14.2 &    11.6 \\
 53264.568 &  0.161 &   -45.1 &    15.7 &    12.6 \\
 53265.491 &  0.184 &   -45.4 &    16.1 &    12.9 \\
 53269.509 &  0.283 &   -54.3 &    17.6 &    13.9 \\
 53271.515 &  0.333 &   -46.1 &    16.4 &    13.2 \\
 53272.541 &  0.358 &   -39.2 &    15.0 &    12.3 \\
 53273.506 &  0.382 &   -30.6 &    13.2 &    10.9 \\
\enddata 
\end{deluxetable} 
 
 \pagebreak

\placetable{tab2}      
\begin{deluxetable}{lcccc} 
\tablewidth{0pc} 
\tablenum{2} 
\tablecaption{Orbital Elements\label{tab2}} 
\tablehead{ 
\colhead{Element} &\colhead{Rivinius et al.\ (2020)}&\colhead{$V_r(1)$}&\colhead{$V_r(2)$}&\colhead{$V_r^{sub}(2)$}} 
\startdata 
$P$ (d)                  \dotfill & $40.333 \pm 0.004$   & $40.334 \pm 0.005$   & \nodata & \nodata \\ 
$T$ (HJD -- 2,400,000)   \dotfill & $53177.44 \pm 0.07$ & $53217.7 \pm 2.2$ & \nodata & \nodata \\ 
$e$                      \dotfill & $0.03 \pm 0.01$      & $0.039 \pm 0.015$   & \nodata & \nodata \\ 
$\omega$ (deg)           \dotfill & 89           & $85 \pm 20$      & \nodata & \nodata \\ 
$K_1$ (km s$^{-1}$)      \dotfill & $61.3 \pm 0.6$     & $61.8 \pm 1.0$   & \nodata & \nodata \\ 
$K_2$ (km s$^{-1}$)      \dotfill & \nodata      & \nodata      & $7.6 \pm 1.0$ & $3.9 \pm 0.7$ \\ 
$V_0$ (km s$^{-1}$)      \dotfill & $9.4 \pm 0.5$      & $7.5 \pm 0.7$      & $7.4 \pm 0.9$ & $7.6 \pm 0.8$ \\ 
$a_1\sin i$ ($R_\odot$)  \dotfill & \nodata      & $49.2 \pm 0.8$     & \nodata & \nodata \\ 
$f(M)$ ($M_\odot$)       \dotfill & $0.96 \pm 0.03$    & $0.99 \pm 0.05$    & \nodata & \nodata \\ 
r.m.s. (km s$^{-1}$)     \dotfill & \nodata      & 4.8          & 1.7     & 2.3     \\ 
\enddata 
\end{deluxetable} 

\begin{figure}[h] 
\begin{center} 
{\includegraphics[angle=90,height=10cm]{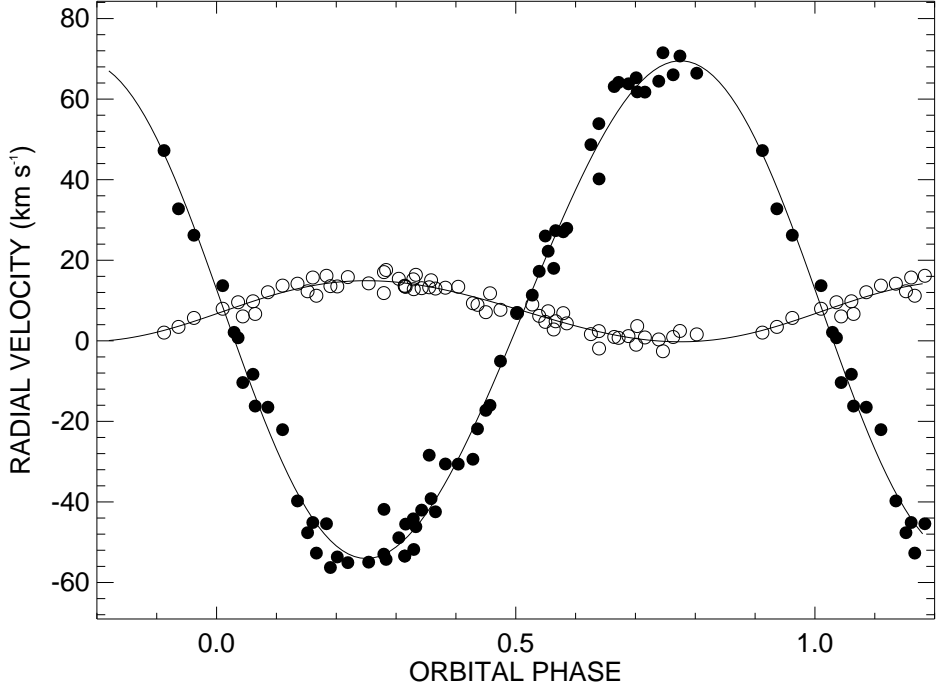}} 
\end{center} 
\caption{The observed and fitted radial velocity curves for 
star 1 (filled circles) and star 2 (open circles).} 
\label{fig1} 
\end{figure}

\placefigure{fig1}     
 
We measured the H$\alpha$ emission line wing bisector velocity 
using the Gaussian sampling method described by \citet{Shafter1986}.
Each observed profile is cross-correlated with a function 
consisting of oppositely-signed Gaussian functions that are 
separated from line center by an amount $dv$, and the 
zero-crossing position of the resulting CCF yields the 
bisector velocity.  We present in column 4 of Table~1 the 
wing bisector velocities $V_r(2)$ that represent the radial 
velocity of the Be star (star 2) that is centered in the disk. 
These velocities are plotted as open circles in Figure~1, 
and they display the orbital anti-phase motion expected for
the companion in the 40~d orbit.  A restricted orbital fit 
of the wing velocities was made by fixing all the elements 
except for the systemic velocity $V_0$ and semiampltude $K_2$, 
and the results are listed in column 4 of Table 2.  The derived value 
of $K_2$ was obtained from velocities measured using Gaussian offsets of 
$dv = \pm 200$ km~s$^{-1}$ (bisector at about $25\%$ of peak strength)
and Gaussian FWHM = 100 km~s$^{-1}$, and the uncertainty 
in $K_2$ reflects the results made with a range of Gaussian 
separation $dv$.  

The spectrum of star 1 probably contains a relatively strong 
H$\alpha$ absorption component that is blended with the emission 
component from the Be star disk.  The absorption wings of the
star 1 component will appear Doppler shifted with the orbit, 
causing the emission wings to be slightly depressed when 
the absorption component moves further into the line wing.
This periodic swing in net emission wing strength will influence
the resulting measurements of the wing bisector.  We modeled 
this line blending process by shifting the model absorption
line to the calculated radial velocity for each spectrum 
and re-scaling the line depth assuming that star 1 contributes
$45\%$ of the total monochromatic flux.
We again used a model from the BLUERED grid \citep{Bertone2008} 
that was broadened for a projected rotational velocity of 
$V\sin i = 48$ km~s$^{-1}$ \citep{Zorec2016}.   
The shifted and rescaled model was then subtracted from 
the observed spectrum to create a version corresponding to 
the Be star and disk alone (see Fig.~3 below).  The wing 
bisector velocities were measured in the same way for the 
model-subtracted spectra, and these are listed in column 5
of Table~1 under the heading $V_r^{sub}(2)$.   An orbital 
fit of these velocities for the model-subtracted profiles 
does indeed yield a smaller but non-zero value of semiamplitude
$K_2$ (column 5 of Table~2).  
We found that there is an approximate linear relationship
between the adopted monochromatic flux scale factor $r=f_1/(f_1+f_2)$ 
and the derived semiamplitude, $K_2 \approx (7.5 - 7.9 ~r)$ km~s$^{-1}$. 
Consequently, in order to reduce the reflex motion to zero, 
we would need to assume a large flux contribution from 
star 1, $r=0.96$.  However, this is far outside of the range 
of possibility based on the appearance of the 
absorption lines of the Be star that indicate that 
it (star 2) is a major flux contributor \citep{Rivinius2020}. 

The wing bisector radial velocity measurements from both 
the observed and model-subtracted spectra indicate that the 
Be disk emission shows the Doppler shifts expected for the 
companion star in the 40~d orbit.  This is compelling evidence 
that the Be star, and not a black hole, is the orbiting 
companion of star 1.  \citet{Rivinius2020} argued that the 
Be star lines are stationary, but the small value of $K_2$ 
makes it very difficult to discern the orbital motion of
the Be star in the features plotted in Figures 2 and C.2
of their paper. 


 
\section{Orbital-Phase Variations in H$\alpha$ Morphology}  

Be disks are dynamic entities that change on a range of 
timescales \citep{Rivinius2013}, but we would only expect to 
observe emission line variations related to orbital phase if
the Be star was part of the binary system.   The variations 
of the H$\alpha$ emission profile as a function of orbital 
phase are presented in Figures 2 and 3 for the cases of 
the directly observed profile and the profiles after 
subtraction of the shifted and scaled model absorption spectrum 
for star 1, respectively.  The upper panels show plots 
of the profiles offset in position according to orbital 
phase while the lower plots show a gray-scale image representation
of the profiles as a function of radial velocity and orbital phase
(from the time of periastron in the orbital fit given in Table 2). 
We found that the emission equivalent width was about $38\%$
stronger in the Set~A observations from 1999 compared to 
those in Set~B from 2004, so we down-scaled the emission 
strength for the 12 spectra from Set~A in order to make the 
comparison with the Set~B profiles in Figures 2 and 3. 

Both figures show the central violet ($V$) and red ($R$) 
peaks that appear approximately constant in radial velocity. 
The half-separation between the peaks is 60 km~s$^{-1}$, 
which matches the semiamplitude of star 1 ($K_1$ in Table 2). 
Simple models for Keplerian disks take the half-separation 
of the peaks as a measure of the projected orbital velocity 
at the outer boundary of the disk, and the fact that this 
is the same as $K_1$ suggests that the outer radius of the 
disk is comparable in size to the separation between 
the stars in the binary if star 1 has a small mass 
(see Section 4).  We would generally expect that the 
disk outer boundary falls well within the Roche lobe of 
the Be star, and thus the emission peak half-separation 
would be larger than $K_1$ (corresponding to Keplerian 
motion with smaller radius).  However, in cases where the 
disk has high gas density and the companion has low mass, 
the disk may extend beyond the companion into a 
circumbinary configuration (see the case of HR 2142; 
\citealt{Peters2016}).  Thus, the similarity of the emission 
peak half-separation and $K_1$ may indicate that the 
disk in HR~6819 extends to or beyond the companion. 

Figure~3 shows the net emission after subtraction of the 
blended absorption feature from star 1, and the upper 
panel shows that there are phase-related variations in the 
$V/R$ peak ratio.  Near phase 0.25 when star 1 is blueshifted, 
$V/R>1$, while the reverse is observed at phase 0.75. 
This implies that the disk density has an azimuthal 
asymmetry with greater emission (larger gas density) in the region
of the disk facing the companion. 

Both Figures 2 and 3 show evidence of small emission enhancements 
in the lines wings that alternate in the blue and red wings 
twice per orbit.  These generally appear near an absolute 
radial velocity of $\approx 100$ km~s$^{-1}$ at the orbital phases
of the radial velocity extrema, and the enhancements move to lower 
velocity as time progresses.  These kinds of moving features 
are also observed in the Be binary HD~55606, and they are interpreted
as the result of a two-arm spiral density wave in the disk 
caused by the tidal force of the companion \citep{Chojnowski2018}.
The inner part of the spiral arm appears first at higher velocity, 
and as the orbit progresses the outer parts of the trailing arms 
create Doppler shifts at the lower Keplerian velocities found
at larger disk radius.  

The presence of orbital-phase related variations in the 
emission line shape can only be explained if the Be star 
is part of the binary system and the changes are related
to the orbital Doppler shifts and binary-induced disk 
structure.  

\begin{figure}[h!] 
\begin{center} 
{\includegraphics[angle=0,height=21cm]{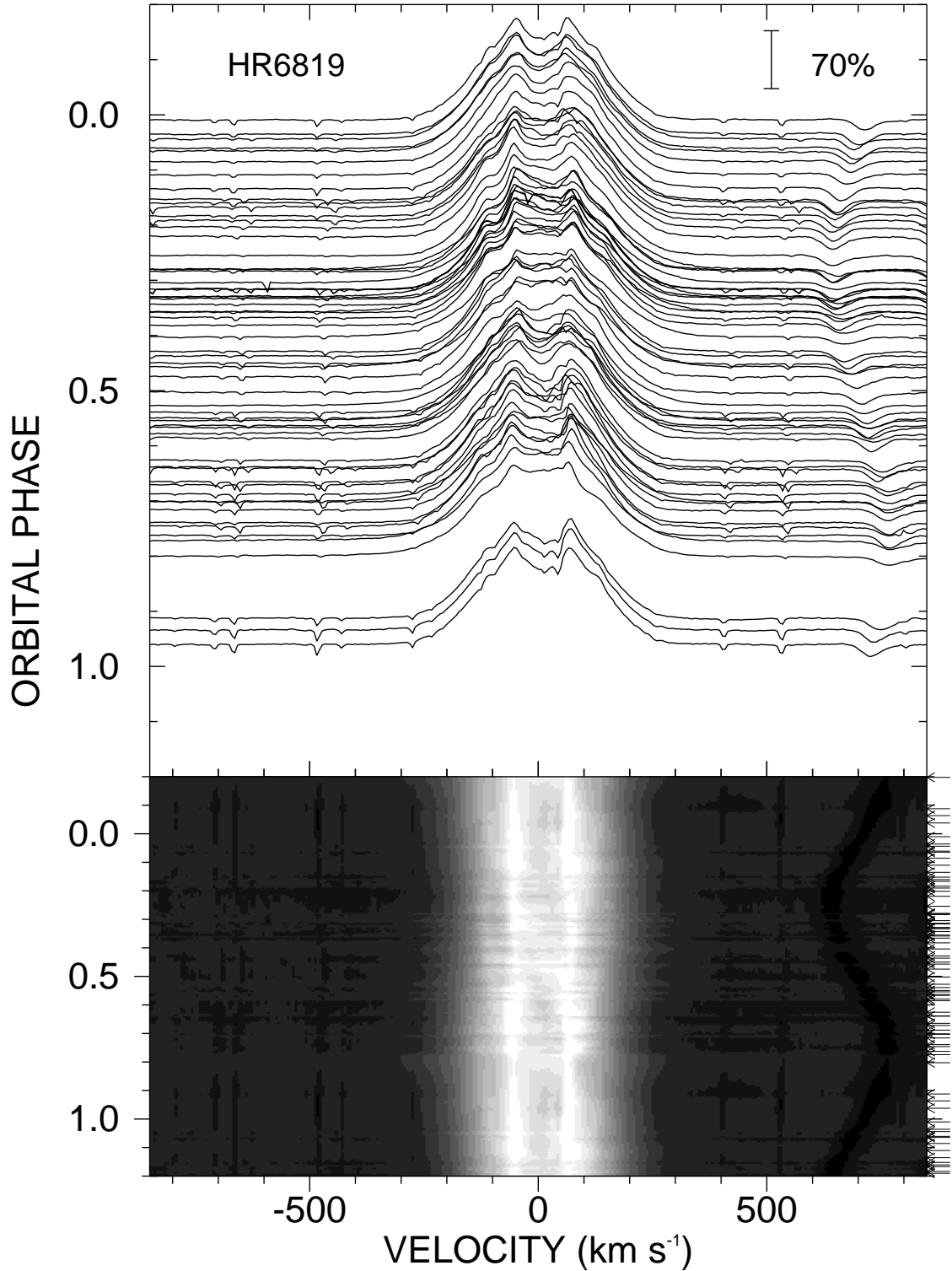}} 
\end{center} 
\caption{The observed H$\alpha$ emission line profiles for HR~6819. 
The top panel shows the individual profiles offset so that the 
continuum is placed at the orbital phase of observation. The lower
panel shows the same profiles transformed into an image as a 
function of radial velocity and orbital phase. The absorption 
component at right is the \ion{C}{2} $\lambda 6578$ feature 
in the spectrum of the B3~III star.} 
\label{fig2} 
\end{figure}

\placefigure{fig2}     
 
\begin{figure}[h!] 
\begin{center} 
{\includegraphics[angle=0,height=21cm]{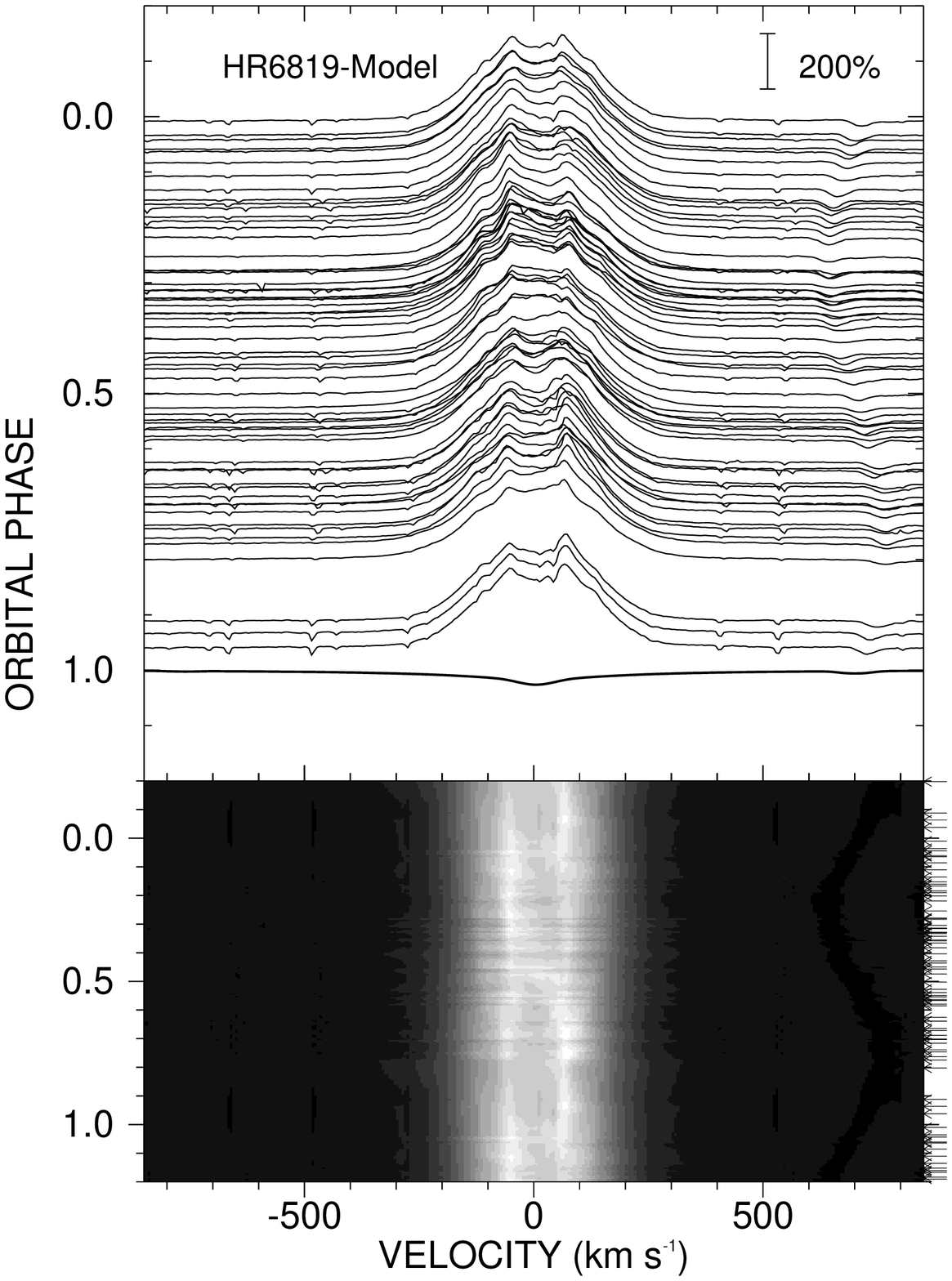}} 
\end{center} 
\caption{The H$\alpha$ emission profiles after subtraction of a 
model profile for the absorption component of the B3~III star assuming  
that it contributes $55\%$ of the monochromatic flux \citep{Rivinius2020}
(shown at the bottom of the top panel without Doppler shift).} 
\label{fig3} 
\end{figure}

\placefigure{fig3}     
 
\section{Binary Model for HR~6819}                  

\citet{Rivinius2020} examined the system masses by assuming a mass
for the B3~III (star 1) component.  Given the evidence above that
the Be star is the more massive component of the 40~d binary, 
it is useful to discuss the system properties by considering 
a probable mass of the Be star.  \citet{Rivinius2020} suggest 
that the Be star is slightly hotter and has a higher gravity 
than the B3~III star, so we assume that Be star has properties 
associated with a B2.5~V star: $M_2/M_\odot = 6$ and $R_2/R_\odot = 3.5$
from the spectral calibration of \citet{Pecaut2013}. 
The mass ratio derived from the radial velocities (Table~2) 
is $M_1/M_2=K_2/K_1=0.06$ for the model-subtracted case and 
0.12 for the directly observed case.  Then the implied 
orbital inclination from the orbital mass function and 
mass ratio is $i\approx 35^\circ$.  This relatively low inclination
is consistent with the moderate value of the Be star projected
rotational velocity $V \sin i \approx 180$ km~s$^{-1}$
(from the broad \ion{He}{1} line components illustrated in 
Fig.~2 and C.2 in \citealt{Rivinius2020}), so that the actual 
equatorial velocity is large, $(V\sin i) / \sin i \approx 310$ 
km~s$^{-1}$, similar to that found in many Be stars. 

The mass of the B3~III star is in the range $0.4 M_\odot$ 
(model-subtracted case) to $0.8 M_\odot$ (directly observed),
much lower than is typical for stars of this classification. 
\citet{Rivinius2020} estimate that the star has an 
effective temperature $T_{\rm eff}=17$~kK, luminosity 
$\log L/L_\odot = 3.5$, and radius $R/R_\odot= 6$. 
Based upon the adopted inclination, $i=35^\circ$, the 
semimajor axis is $a=93 R_\odot$ and the stellar radii of 
both stars are much smaller than their respective Roche 
radii. 

We suggest that the B3~III star has a low mass because it 
is the stripped down remains of what was originally the 
mass donor prior to the completion of mass transfer. 
\citet{Istrate2016} present models for the evolution of 
such precursors to He white dwarf stars, and we show in 
Figure~4 the evolutionary tracks for the two largest 
core remnant masses in their basic grid for solar metallicity 
stars.  The cores initially grow in temperature at nearly 
constant luminosity before fading as they approach the 
white dwarf cooling sequence.  In some cases, the models 
indicate that the cores will experience episodic H-shell 
burning episodes that will lead to periods of increased 
luminosity.  We also show the location of the B3~III star in 
HR~6819 at the top of the $(\log T_{\rm eff}, \log L/L_\odot)$ 
diagram.  By extrapolation to the higher luminosity of 
the B3~III star, the evolutionary models appear to indicate 
a mass of $M_1/M_\odot = 0.44$ that is consistent with the 
estimates given above.  We note that the implied gravity 
from the mass and radius is $\log g = 2.5$, which is smaller
than that derived by \citet{Rivinius2020} of $\log g = 3.5$
from spectroscopic diagnostics. However, the radius of 
the B3~III star may need revision once the relative flux 
contribution of the star is better determined. 

\begin{figure}[h!] 
\begin{center} 
{\includegraphics[angle=90,height=10cm]{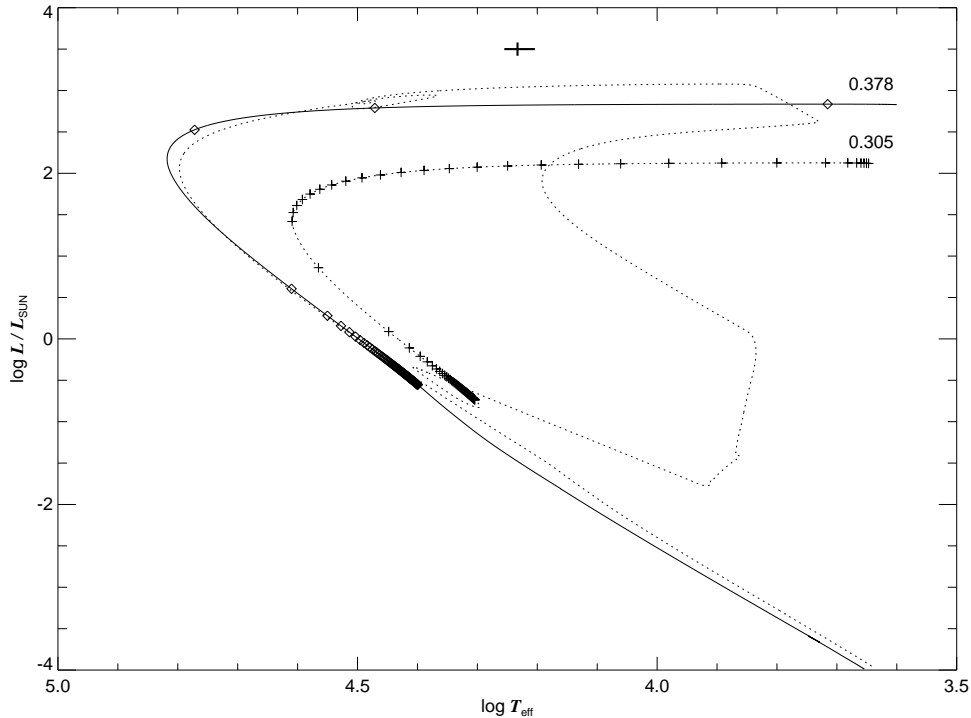}} 
\end{center} 
\caption{Evolutionary tracks in the $(\log T_{\rm eff}, \log L/L_\odot)$ 
plane for models of a $0.305 M_\odot$ (dotted line) and 
a $0.378 M_\odot$ (solid line) stripped core from 
the work of \citet{Istrate2016}. 
Diamonds and plus signs mark the locations at increments 
of 0.1 Myr from the starting point in the model (upper right) 
to an elapsed time of 10 Myr (center). 
The cross at the top shows the estimated 
parameters for the B3~III star in HR~6819.} 
\label{fig4} 
\end{figure}

\placefigure{fig4}     

The evolutionary tracks in Figure~4 are marked by diamonds
($0.378 M_\odot$) and plus signs ($0.305 M_\odot$) at intervals 
of 0.1 Myr, and the stripped companion in HR~6819 falls near
to the stage in the $0.378 M_\odot$ model where the star 
crosses the diagram at constant luminosity in only $\approx 0.1$~Myr.
This indicates that we observe the system at a young, fleeting, and
uncommon stage in its post-mass transfer evolution. 

Our examination of the H$\alpha$ emission line in the spectrum
of HR~6819 shows that the Be star disk varies in Doppler shift
and line shape with the orbital phase.  Consequently, the 
underlying Be star is the probable companion of the low mass 
B3~III in the 40~d orbit.  The HR~6819 binary shares many of the 
characteristics found in other Be + He core systems (in particular, 
mass ratio and orbital period), but the companions are usually much 
fainter and hotter than the Be host stars \citep{Wang2018}.  The luminous 
and low mass companion in the HR~6819 system may represent 
a rare and important case in which the companion has recently 
completed mass transfer and has yet to descend to the white 
dwarf cooling stage of evolution. 

{\it Note Added in Manuscript:} After this work was completed, we became 
aware of two other papers that come to essentially the same conclusion.
\citet{Bodensteiner2020} and \citet{ElBadry2020} performed spectral 
disentangling to reconstruct the spectral lines of both components 
from the composite spectra.  Both studies explored spectral reconstructions
over a range in test $K_2$ values, and they found that the 
$\chi^2$ residuals between the observed and model binary spectra
obtained a minimum for $K_2 = 4.0 \pm 0.8$ km~s$^{-1}$ 
and $K_2 = 4.5 \pm 2$ km~s$^{-1}$, respectively. 
These agree within errors with our result $K_2 = 3.9 \pm 0.7$ km~s$^{-1}$
after subtraction of a model H$\alpha$ absorption profile scaled to 
a $45\%$ flux contribution from the B3~III star. 

 
\acknowledgments 
 
We thank an anonymous referee for their very helpful comments.
This material is based upon work supported by the 
National Science Foundation under Grant No.~AST-1908026. 
Institutional support has been provided from the GSU College 
of Arts and Sciences.  We are grateful for all this support. 
 
 
\facilities{ESO:1.52m, Max Planck:2.2m, IUE}



\bibliographystyle{aasjournal}
\bibliography{paper}{}

\begin{thebibliography}{}
\expandafter\ifx\csname natexlab\endcsname\relax\def\natexlab#1{#1}\fi
\providecommand{\url}[1]{\href{#1}{#1}}
\providecommand{\dodoi}[1]{doi:~\href{http://doi.org/#1}{\nolinkurl{#1}}}
\providecommand{\doeprint}[1]{\href{http://ascl.net/#1}{\nolinkurl{http://ascl.net/#1}}}
\providecommand{\doarXiv}[1]{\href{https://arxiv.org/abs/#1}{\nolinkurl{https://arxiv.org/abs/#1}}}

\bibitem[{{Bertone} {et~al.}(2008){Bertone}, {Buzzoni}, {Ch{\'a}vez}, \&
  {Rodr{\'\i}guez-Merino}}]{Bertone2008}
{Bertone}, E., {Buzzoni}, A., {Ch{\'a}vez}, M., \& {Rodr{\'\i}guez-Merino},
  L.~H. 2008, \aap, 485, 823, \dodoi{10.1051/0004-6361:20078923}

\bibitem[{{Bodensteiner} {et~al.}(2020){Bodensteiner}, {Shenar}, {Mahy},
  {Fabry}, {Marchant}, {Abdul-Masih}, {Banyard}, {Bowman}, {Dsilva}, {Frost},
  {Hawcroft}, {Reggiani}, \& {Sana}}]{Bodensteiner2020}
{Bodensteiner}, J., {Shenar}, T., {Mahy}, L., {et~al.} 2020, A\&A, submitted,
  arXiv:2006.10770.
\newblock \doarXiv{2006.10770}

\bibitem[{{Chojnowski} {et~al.}(2018){Chojnowski}, {Labadie-Bartz}, {Rivinius},
  {Gies}, {Panoglou}, {Borges Fernandes}, {Wisniewski}, {Whelan}, {Mennickent},
  {McMillan}, {Dembicky}, {Gray}, {Rudyk}, {Stringfellow}, {Lester},
  {Hasselquist}, {Zharikov}, {Levenhagen}, {Souza}, {Leister}, {Stassun},
  {Siverd}, \& {Majewski}}]{Chojnowski2018}
{Chojnowski}, S.~D., {Labadie-Bartz}, J., {Rivinius}, T., {et~al.} 2018, \apj,
  865, 76, \dodoi{10.3847/1538-4357/aad964}

\bibitem[{{El-Badry} \& {Quataert}(2020)}]{ElBadry2020}
{El-Badry}, K., \& {Quataert}, E. 2020, MNRAS, submitted, arXiv:2006.11974.
\newblock \doarXiv{2006.11974}

\bibitem[{{Istrate} {et~al.}(2016){Istrate}, {Marchant}, {Tauris}, {Langer},
  {Stancliffe}, \& {Grassitelli}}]{Istrate2016}
{Istrate}, A.~G., {Marchant}, P., {Tauris}, T.~M., {et~al.} 2016, \aap, 595,
  A35, \dodoi{10.1051/0004-6361/201628874}

\bibitem[{{Morbey} \& {Brosterhus}(1974)}]{Morbey1974}
{Morbey}, C.~L., \& {Brosterhus}, E.~B. 1974, \pasp, 86, 455,
  \dodoi{10.1086/129630}

\bibitem[{{Pecaut} \& {Mamajek}(2013)}]{Pecaut2013}
{Pecaut}, M.~J., \& {Mamajek}, E.~E. 2013, \apjs, 208, 9,
  \dodoi{10.1088/0067-0049/208/1/9}

\bibitem[{{Peters} {et~al.}(2016){Peters}, {Wang}, {Gies}, \&
  {Grundstrom}}]{Peters2016}
{Peters}, G.~J., {Wang}, L., {Gies}, D.~R., \& {Grundstrom}, E.~D. 2016, \apj,
  828, 47, \dodoi{10.3847/0004-637X/828/1/47}

\bibitem[{{Pols} {et~al.}(1991){Pols}, {Cote}, {Waters}, \& {Heise}}]{Pols1991}
{Pols}, O.~R., {Cote}, J., {Waters}, L.~B.~F.~M., \& {Heise}, J. 1991, \aap,
  241, 419

\bibitem[{{Rivinius} {et~al.}(2020){Rivinius}, {Baade}, {Hadrava}, {Heida}, \&
  {Klement}}]{Rivinius2020}
{Rivinius}, T., {Baade}, D., {Hadrava}, P., {Heida}, M., \& {Klement}, R. 2020,
  \aap, 637, L3, \dodoi{10.1051/0004-6361/202038020}

\bibitem[{{Rivinius} {et~al.}(2013){Rivinius}, {Carciofi}, \&
  {Martayan}}]{Rivinius2013}
{Rivinius}, T., {Carciofi}, A.~C., \& {Martayan}, C. 2013, \aapr, 21, 69,
  \dodoi{10.1007/s00159-013-0069-0}

\bibitem[{{Shafter} {et~al.}(1986){Shafter}, {Szkody}, \&
  {Thorstensen}}]{Shafter1986}
{Shafter}, A.~W., {Szkody}, P., \& {Thorstensen}, J.~R. 1986, \apj, 308, 765,
  \dodoi{10.1086/164549}

\bibitem[{{Sigut} {et~al.}(2020){Sigut}, {Mahjour}, \& {Tycner}}]{Sigut2020}
{Sigut}, T.~A.~A., {Mahjour}, A.~K., \& {Tycner}, C. 2020, \apj, 894, 18,
  \dodoi{10.3847/1538-4357/ab8386}

\bibitem[{{Wang} {et~al.}(2018){Wang}, {Gies}, \& {Peters}}]{Wang2018}
{Wang}, L., {Gies}, D.~R., \& {Peters}, G.~J. 2018, \apj, 853, 156,
  \dodoi{10.3847/1538-4357/aaa4b8}

\bibitem[{{Zorec} {et~al.}(2016){Zorec}, {Fr{\'e}mat}, {Domiciano de Souza},
  {Royer}, {Cidale}, {Hubert}, {Semaan}, {Martayan}, {Cochetti}, {Arias},
  {Aidelman}, \& {Stee}}]{Zorec2016}
{Zorec}, J., {Fr{\'e}mat}, Y., {Domiciano de Souza}, A., {et~al.} 2016, \aap,
  595, A132, \dodoi{10.1051/0004-6361/201628760}

\end{thebibliography}
 
\end{document}